\documentclass[twocolumn,showpacs,preprintnumbers,amsmath,amssymb]{revtex4}
\usepackage{graphicx}
\usepackage{dcolumn}
\usepackage{bm}
\def\bea {\begin{eqnarray}}
\def\eea {\end{eqnarray}}

\def\be {\begin{equation}}
\def\ee {\end{equation}}

\begin{document}
\title
{Radial Flow from Electromagnetic Probes and Signal of Quark Gluon Plasma
}

\author{Payal Mohanty, Jajati K Nayak, Jan-e Alam and Santosh K Das}
\medskip
\affiliation{Variable Energy Cyclotron Centre, 1/AF, Bidhan Nagar,
Kolkata - 700064, INDIA
}
\begin{abstract}
An attempt has been made to extract the evolution of radial flow
from the analysis of the experimental data on electromagnetic probes
measured at SPS and RHIC energies.  The transverse momentum ($p_T$) spectra
of photons and dileptons measured by WA98 and NA60 collaborations
respectively at the CERN-SPS and the photon and dilepton 
spectra  obtained by PHENIX collaboration at BNL-RHIC have been used to  
constrain the theoretical models.  We use the ratio of photon 
to dilepton spectra to extract the flow,  where some model dependence 
are canceled out.  Within the ambit of the present analysis we argue that 
the variation of the radial velocity with invariant mass is indicative of a 
phase transition from initially produced partons to hadrons at SPS and RHIC 
energies.
\end{abstract}

\maketitle

\section{Introduction}
The hot and dense matter expected to be 
formed in the partonic phase after ultra-relativistic
heavy ion collisions
dynamically evolve in space and time due to high internal pressure.
Consequently the system cools and reverts back to hadronic matter from the
partonic phase.  Just after the formation,
the entire energy of the system is thermal in nature
and with progress of time some
part of the thermal energy gets converted to the collective (flow) energy.
In other words, during the expansion stage
the total energy of the system is shared by the thermal
as well as the collective degrees of freedom. The evolution of
the collectivity within the system is sensitive to the Equation of State (EoS).
Therefore, the study of the collectivity in the
system formed in the quark gluon plasma (QGP) phase will be useful to
shed light on the EoS~\cite{heinz,hung,hirano} (see~\cite{pasi,teaney} 
for review) and on the nature
of the transition that may take place during the 
evolution process.  It is well known that the average magnitude of
radial flow at the freeze-out surface can be extracted from
the transverse momentum ($p_T$) spectra of the hadrons.
However, hadrons being strongly interacting objects
can bring the information of the state of the system
when it is too dilute to support collectivity {\it i.e.}
the parameters of collectivity extracted from the hadronic
spectra are limited to the evolution stage where the 
collectivity ceases to exist. These collective parameters have 
hardly any information about the interior of the matter. 
On the other hand electromagnetic (EM) probes, {\it i.e.}
photons and dileptons are produced and emitted~\cite{mclerran,gale,weldon}
(see~\cite{alam1,alam2,rapp} for review) from each space time points. 
Therefore, estimating radial flow from the EM probes will shed light on 
the time evolution of the collectivity in the system~\cite{renk}.

The invariant momentum distribution of
photons produced from a thermal source depends on the temperature ($T$)
of the source through the thermal phase space 
distributions of the participants of the
reaction that produces the photon~\cite{cywong}.
As a result the $p_T$ spectra of photon reflects the temperature
of the source.  Hence ideally the photons 
with intermediate $p_T$ values ($\sim 2 - 3$ GeV,
depending on the value of initial temperature) 
reflect the properties of QGP (realized when $T > T_c$, $T_c$ is 
the transition temperature). Therefore, one should look into the 
$p_T$ spectra for these values of $p_T$ for the detection of QGP. 
However, for an expanding system 
the situation is far more complex. The thermal phase space factor changes
 - several factors {\it e.g.} the transverse
kick received by  low $p_T$ photons due to flow originating from the low
temperature hadronic phase (realized when $T<T_c$) populates
the high $p_T$ part of the spectra~\cite{ja1993}.
As a consequence the intermediate or the high $p_T$
part of the spectra  contains contributions from both QGP and hadrons.
For dilepton the situation is, however, different because in this
case we have two kinematic variables - out of these two, the
$p_T$ spectra is affected by the flow,
however, the $p_T$ integrated invariant mass ($M$) spectra
is unaltered by the  flow in the system.  It should be mentioned 
here that for $M$ below $\rho$ peak and  above $\phi$ peak dileptons
from QGP dominates over its hadronic counterpart (assuming the contributions
from hadronic cocktails are subtracted out) within the framework of 
the present model. However, the spectral function of low mass vector mesons 
(mainly $\rho$) may shift toward lower invariant mass region 
due to non-zero temperature and density effects. As a consequence 
the contributions from the decays of such vector mesons to lepton 
pairs could populate the low $M$ window and may dominate over the
contributions from the QGP phase~\cite{rgrlk} (and also ~\cite{rapp,alam2}
for review).  In the present work such thermal effects are not considered. 
All these suggests that a judicious choice of $p_T$ and 
$M$ windows will be very useful to characterize the
flow in QGP and hadronic phase. However, there are still some
difficulties.  The calculations of EM probes from thermal sources  
depend on the parameters like initial 
temperature ($T_i$), thermalization time ($\tau_i$),  
chemical freeze-out temperature ($T_{\mathrm ch}$), 
kinetic freeze-out temperature ($T_f$) etc, which are not known uniquely. 
To minimize the dependence of thermal sources on these parameters 
the importance of the ratio of the transverse momentum spectra of 
photon to dilepton has been emphasized previously~\cite{sinha,JKN1,JKN2}
in order to overcome the above mentioned uncertainties.
It may be mentioned here that in the limit of $M\rightarrow 0$
the lepton pairs (virtual photons) emerge as real photons. 
Therefore, the evaluation of the ratio of the $p_T$ spectra of photons to 
dileptons for  various invariant mass bins along with 
a judicious choice of the $p_T$ and $M$ windows 
will be very useful to extract the properties of QGP as well as that of
hadronic phase. This will be demonstrated in the present work by analyzing
WA98 and PHENIX photons and NA60 and PHENIX dilepton spectra. 

The paper is organized as follows. In the next section,  
the production of thermal  photons and dileptons
are briefly outlined.  In section III
the expansion dynamics of the system is discussed.  
Section IV is devoted
to results and discussions and finally in section V, we present 
summary and conclusions. 
\section{Photons and dileptons production}
The $p_T$ and $M$ spectra of EM probes 
measured in the experiments are mingled with all the sources of 
productions -  broadly categorized as: (i) prompt productions resulting
from the interactions of the partons of the colliding  nuclei, 
(ii) thermal productions from the interactions of thermal partons as well as
from thermal hadrons and (iii) finally from the decays of the long 
(compared to strong interaction time scale) lived mesons.  The contributions
from $pp$ collisions at a given collision energy 
can be used as a bench-mark to estimate the hard contributions.
To estimate the thermal contributions we adopt the following procedure:
{\it thermal contributions=contributions from heavy ion collision minus
$N_{\mathrm coll}\times $contributions from $pp$ collisions} where
$N_{\mathrm coll}$ is the number of nucleon-nucleon interactions in
the nuclear collisions at a given centrality. Instead of evaluating
the hard contributions by applying pQCD we use the
experimental data both for photons and dileptons 
from the $pp$ collisions wherever available to  minimize the uncertainties
in the  contributions of category (i).

For collisions with large nuclei {\it e.g.} $Au+Au$ or $Pb+Pb$,
the number of valence $d$ quarks are more than the number of valence $u$ quark
because in these nuclei there are more neutrons than protons.
The magnitude of electric charge of the $d$ quark is half of that of 
$u$ quarks.
Consequently, in the production of EM probes from $Au+Au$ collisions - 
a larger number $d$ quarks with lesser charge and comparatively 
a smaller number of $u$ quarks with larger charge are involved. 
Therefore, for the photon production from  $Au+Au$ interaction 
for the category (i) is  not a mere superposition of
the yield from $p+p$ interaction. However,  in the present 
work we concentrate in the kinematic region of central rapidity
where the number of valence quarks are small. In fact, this
is negligible for  RHIC energy.
It is worth mentioning here that the role of EM probes produced from
other mechanisms - 
like fragmentation of high energy quarks and the interaction of high
energy partons with thermal QGP medium~\cite{tgfh} are ignored here.
It is expected that at the $p_T$ domain of our interest
omission of these mechanism will not change the final results
significantly.

\subsection{Production of thermal photons and dileptons}

For the present work photons and dileptons from thermalized matter
of partons and hadrons play the most crucial role.
The rate of thermal dilepton production per unit space-time volume
per unit four momentum volume
is given by~\cite{mclerran,gale,weldon,alam1}
\begin{equation}
\frac{dR}{d^4p}=\frac{\alpha}{12\pi^4 p^2}L(p^2){\mathrm Im}\Pi_{\mu}^{R\mu}
f_{BE}
\label{eq2}
\end{equation}
where $\alpha$ is the EM coupling constant, ${\mathrm Im}\Pi_{\mu}^{\mu}$ is the
imaginary part of the retarded photon self energy and $f_{BE}(E,T)$ is the
thermal phase space distribution for Bosons. 
$L(p^2)$=$(1+2m^2/p^2)\sqrt{1-4m^2/p^2}$, arises
from the final state leptonic current involving Dirac spinors of mass $m$.
As mentioned before, in the limit of vanishing $M$ a lepton pair
{\it i.e.} a virtual photon appears as a  real photon. Therefore, 
the real photon production rate can be obtained from the dilepton
emission rate by replacing the product of EM vertex
$\gamma^{\star}$ $\rightarrow$ $ l^{+}l^{-}$, the term involving
final state leptonic current and the square of the (virtual) photon
propagator by the polarization sum for the real photon.
For an expanding system
$E$ should be replaced by $u_\mu p^\mu$, where $p^\mu$ and $u^\mu$ are the four
momentum and the hydrodynamic four velocity respectively.

\subsection{Thermal photons}
The Hard Thermal Loop~\cite{braaten} approximations has been used by several
authors~\cite{photons} to evaluate the photon spectra originating from a 
thermal source of quarks and gluons. 
The complete calculation of emission rate of photons from QGP to order
O$(\alpha\alpha_s)$ has been done
by resuming ladder diagrams in the effective theory~\cite{arnold}, which
has been used in the present work. A set of  hadronic reactions
with all possible isospin combinations have been considered for the 
production of
photons~\cite{npa1,npa2,turbide} from hadronic matter.
The effect of hadronic dipole
form factors has been taken into account in the present
work as in~\cite{turbide}.

\subsection{Thermal dileptons}
The lowest order process producing lepton pairs is $q$ and $\bar{q}$
annihilation. The correction
of order $\alpha_s\alpha^2$ to
the lowest order rate of dilepton production from QGP 
has been calculated in
~\cite{altherr,thoma}, which is considered in the present work.
For the low $M$ dilepton production from the hadronic phase the
decays of the light vector mesons $\rho, \omega$ and $\phi$ has 
been considered in ~\cite{alam2,rapp,shuryak}. 
The continuum part of the vector mesons spectral functions 
constrained by experimental data~\cite{shuryak} have
been included here.
As mentioned before the contributions from the  QGP phase dominates
the $M$ spectra of the lepton pairs below $\rho$-peak
and above the $\phi$-peak if no thermal effects of the
spectral functions
of the vector mesons~(see \cite{alam2,rapp,BR} for review) are considered.
Since the continuum part of the vector meson spectral functions are included
in the current work the processes like four pions annihilations~\cite{4pi}
are excluded
to avoid double counting. 

\begin{figure}[h]
\begin{center}
\includegraphics[scale=0.45]{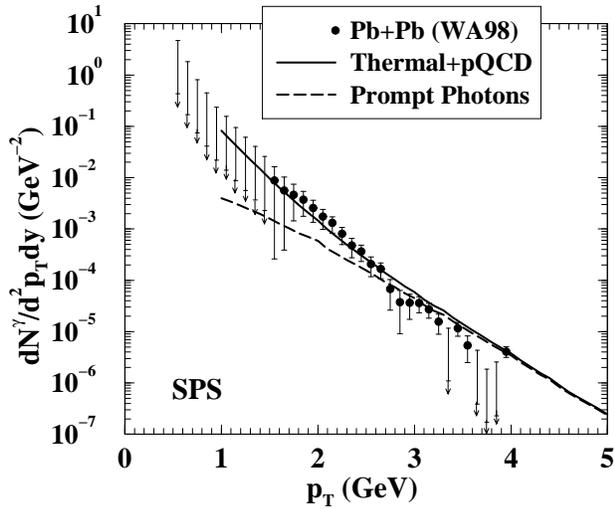}
\caption{Transverse momentum spectra of photon at SPS energy for 
Pb+Pb collision  at mid rapidity.}
\label{fig1}
\end{center}
\end{figure} 
\begin{figure}[h]
\begin{center}
\includegraphics[scale=0.45]{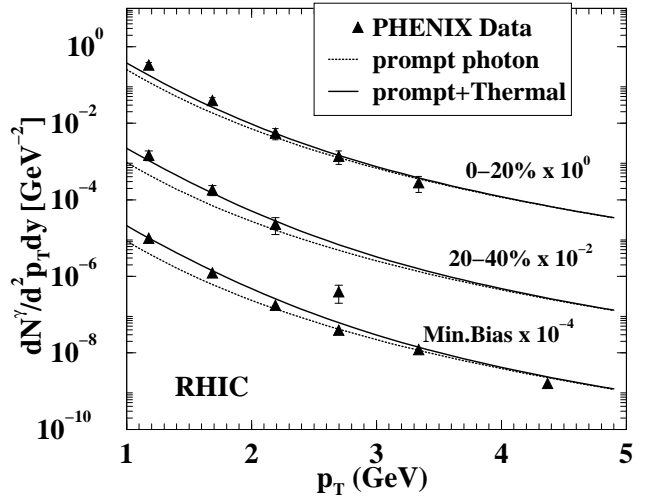}
\caption{Transverse momentum spectra of photons at RHIC energy 
for Au-Au collision for different centralities at mid-rapidity.
}
\label{fig2}
\end{center}
\end{figure} 
\section{Expansion Dynamics}
The space time evolution of the system formed in heavy ion
collisions has been studied by using relativistic hydrodynamics 
with  longitudinal boost invariance~\cite{bjorken} and cylindrical symmetry
~\cite{hvg}.  
We assume that the system reaches equilibration at a time 
$\tau_i$ after the collision. The initial  temperature, 
$T_i$ can be related to the measured hadronic 
multiplicity ($dN/dy$) by the following relation:
\begin{equation}
T_i^{3}\tau_i \approx \frac{2\pi^4}{45\zeta(3)}\frac{1}{4a_{eff}}\frac{1}
{\pi R_A^2}\frac{dN}{dy}.
\label{eq4} 
\end{equation}
where $R_A$ is the radius of the system, $\zeta(3)$  
is the Riemann zeta function and  $a_{eff}=\pi^2g_{eff}/90$ where
$g_{eff}$ ($=2\times 8+ 7\times 2\times 2\times 3\times N_F/8$) is the 
degeneracy of quarks and gluons in QGP, $N_F$=number of flavours. 
The value of $dN/dy$ for various beam energies 
and centralities are calculated from the following equation~\cite{kharzeev}:
\begin{equation}
\frac{dN}{dy}=(1-x)\frac{dn_{pp}}{dy} \frac{<N_{part}>}{2}
+ x\frac{dn_{pp}}{dy}<N_{coll}>
\label{eq5}
\end{equation}
and tabulated in table I. 
\begin{table}[h]
\caption{The values of various parameters - thermalization
time ($\tau_i$), initial temperature ($T_i$) 
and hadronic multiplicity $dN/dy$  - used 
in the present calculations.}
\begin{tabular}{lccccr}
\tableline
$\sqrt{s_{NN}}$ &centrality  & $\frac{dN}{dy}$&$\tau_i(fm)$&$T_i$(MeV) \\
\tableline
17.3 GeV &0-06\%&700&1.0&200\\
200 GeV  &0-20\%&496&0.6&227\\
&20-40\%&226&0.6&203\\
&min. bias&184&0.6&200\\
\tableline
\end{tabular}
\end{table}
$N_{coll}$ is the number of 
collisions and contribute to $x$ fraction to the multiplicity $dn_{pp}/dy$ 
measured in $pp$ collision. The number of participants,
$N_{part}$ contributes 
fraction $(1-x)$ of $dn_{pp}/dy$. The values of $N_{part}$ 
and $N_{coll}$  are estimated by using Glauber Model and
the results are in agreement with~\cite{adler}. We have used
$dn_{pp}/dy=2.43$ and $x=0.1$ at $\sqrt{s_{NN}}=200$ GeV.
It should be mentioned here that the values of $dN/dy$ (through 
$N_{\mathrm part}$ and $N_{\mathrm coll}$ in Eq.~\ref{eq5})
and hence the $T_i$ (through $dN/dy$ in Eq.~\ref{eq4}) depend 
on the centrality of the collisions. The values of $R_A$ for 
different centralities have been evaluated by using the equation 
$R_A\sim 1.1N_{\mathrm part}^{1/3}$.
Some comments are in order here regarding the flow for non-central
collisions which results in non-cylindrical geometrical shape of
the system formed after the collisions. In the present work the non-centrality
of the nuclear collisions is reflected in the initial temperature through
hadronic multiplicity. The geometry of the system due to non-centrality,
should in principle be treated  by (2+1) dimensional~\cite{2plus1d}
or for more rigorous treatment (3+1) dimensional 
hydrodynamical~\cite{3plus1d} evolution.
However, we expect that the results obtained in the present work 
will not be affected substantially due to non-cylindrical  geometric shape
of the system.  Because the flow has been extracted from the ratio of photon
to dilepton spectra for a given centrality. 

We use the EoS obtained from the lattice QCD calculations by the MILC 
collaboration~\cite{MILC}. 
We consider  
kinetic freeze out temperature, $T_f$=140 MeV for all the hadrons. 
The ratios of various hadrons measured experimentally at
different $\sqrt{s_{\mathrm {NN}}}$
indicate that the system formed in heavy ion collisions chemically decouple
at $T_{\mathrm {ch}}$ which is higher than $T_f$ which can be determined 
by the transverse spectra of hadrons~\cite{npa2005}.
Therefore, the system remains out of chemical equilibrium
from $T_{\mathrm {ch}}$ to $T_f$. The deviation of the system from
the chemical equilibrium is taken in to account by introducing 
chemical potential for each hadronic species.
The chemical non-equilibration affects the yields  
through the phase space  factors of the hadrons which in turn
affects the productions of the EM probes.
The value of the chemical potential has been taken in to account
following Ref.~\cite{hirano}. It is expected that the chemical
potentials do not change much for the inclusion of resonances
above $\Delta$.

\section{Results and Discussion}
\subsection{$p_T$ distributions of photons and dileptons}
The prompt photons and dileptons (Drell-Yan) are normally estimated
by using perturbative QCD. 
However, to minimize the theoretical model dependence here
we use the available experimental data
from $pp$ collisions to estimate the hard photon and
dilepton contributions in heavy
ion collisions. 
The WA98 photon spectra from Pb+Pb collisions 
is measured at $\sqrt{s_{NN}}=17.3$ GeV. However,
no data at this collision energy is available for pp interactions. 
Therefore, prompt photons for p+p collision at $\sqrt{s_{NN}}=19.4$ GeV 
has been used~\cite{e704} to estimate the hard contributions for
nuclear collisions at $\sqrt{s_{\mathrm NN}}=17.3$ GeV.
Appropriate  scaling~(see \cite{wa98} for details) 
has been used to obtain the results at $\sqrt{s_{NN}}=17.3$ GeV.  
For the Pb+Pb collisions the result has been appropriately scaled
by the number of collisions at this energy
(this is shown in Fig.~\ref{fig1} as prompt photons).
The high $p_T$ part of the WA98 data is reproduced by the prompt 
contributions reasonably well.  
At low $p_T$ the hard contributions under estimate the data 
indicating the presence of a thermal source.
The thermal photons with initial temperature $=200$ MeV 
along with the prompt contributions explain the WA98 data well 
(Fig.~\ref{fig1}),
with the inclusion of non-zero chemical potentials for 
all hadronic species considered~\cite{hirano}(see also~\cite{renk2}). 
In some of the previous works
~\cite{ashns,dksbs,huovinen,gallmeister,steffen,rupa} the effect of 
chemical freeze-out is ignored. As a result  either a higher value of
$T_i$ or a substantial reduction of hadronic masses in the
medium was required~\cite{ashns}. In the present work, the data has been 
reproduced without any such effects. 

\par
Following a similar procedure, the data~\cite{adare} from Au-Au collisions  at 
$\sqrt{s_{NN}}=200$ GeV has been reproduced well 
by adding the prompt contributions (which is constrained by 
pp data at the same energy) to the thermal photons.
The reproduction of data is satisfactory (Fig.~\ref{fig2}) 
for all the centralities with the initial temperature
shown in table I (see also~\cite{liu}.

\begin{figure}[h]
\begin{center}
\includegraphics[scale=0.45]{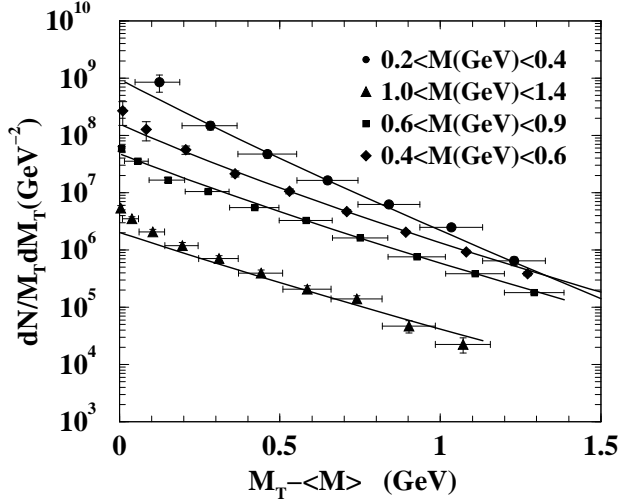}
\caption{Transverse mass spectra of dimuons in In+In 
collisions at SPS energy. Solid lines denote the theoretical results.}
\label{fig3}
\end{center}
\end{figure} 
\begin{figure}[h]
\begin{center}
\includegraphics[scale=0.4]{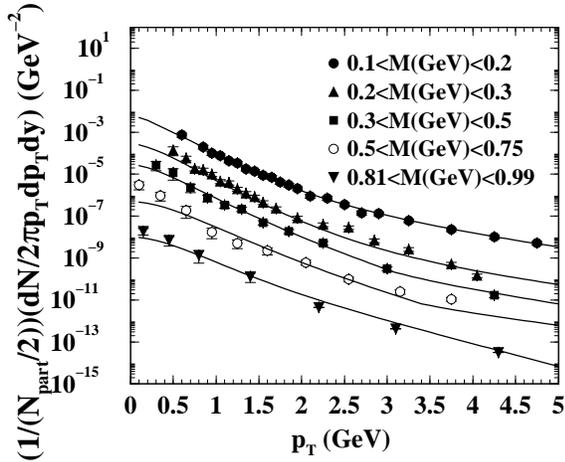}
\caption{Transverse momentum spectra of dileptons for different invariant mass
windows for minimum bias Au-Au collisions at RHIC energy. 
}
\label{fig4}
\end{center}
\end{figure} 
The transverse mass distribution of dimuons produced in In+In collisions
at $\sqrt{s_{NN}}=17.3$ GeV 
has been evaluated for different invariant mass ranges 
(see~\cite{anhs} for details). The quantity $dN/M_TdM_T$ has
been obtained by integrating the production rates over invariant
mass windows $M_{\mathrm min}$ to $M_{\mathrm max}$ and $M_T$ is
defined as $\sqrt{<M>^2+p_T^2}$ where $<M>=(M_{\mathrm min}+M_{\mathrm max})/2$.
The results are compared with the data obtained by NA60 
collaborations~\cite{NA60} at SPS energy (Fig.~\ref{fig3}). 
Theoretical results contain contributions from the 
thermal decays of light vector mesons ($\rho$, $\omega$ and $\phi$) 
and also from the decays of vector mesons at the freeze-out~\cite{cooper}
of the system has also  been considered.  
The non-monotonic variation of the effective slope parameter extracted from
the $M_T$ spectra of the lepton pair with $\langle M\rangle$ 
evaluated within the ambit 
of the present model ~\cite{anhs} 
reproduces the NA60~\cite{NA60} results reasonably well.  
For Au+Au collisions at $\sqrt{s_{NN}}$=200 GeV,
we have evaluated 
the dilepton spectra for different invariant mass bins with the
initial condition (min bias) shown in Table I and lattice QCD equation 
of state. The results are displayed in Fig.~\ref{fig4}.
The  slopes of the experimental data 
on $p_T$ distribution of lepton pairs for different invariant mass windows
measured by the PHENIX collaboration~\cite{phenixlepton} could  be reproduced
well with the same initial condition that reproduces photon 
spectra~\cite{adare}. In fact, the reproduction of data
for the higher mass windows $0.5<$M(GeV)$<0.75$ and $0.81<$M(GeV)$<0.99$ 
do not need any normalization factors (Fig.~\ref{fig4}). 
For lower mass windows slopes are reproduced well but fail to reproduce
the absolute normalization. Therefore, it should be clarified here that
the theoretical results shown in Fig.~\ref{fig4} for lower mass 
windows (to be precise for $0.1<$M(GeV)$<0.2$, $0.2<$M(GeV)$<0.3$ 
and $0.3<$M(GeV)$<0.5$) contain arbitrary normalization constant.
However, it should also be mentioned at this point that 
for the extraction of the flow within the present approach the 
absolute normalization is not essential,  which is
essential is the slope (Eq.~\ref{radvel}).
Therefore, the non-reproduction of the
absolute normalization of the $p_T$ spectra of lepton
pairs for the lower mass windows does not affect the 
extraction of the magnitude of the radial flow. 
\begin{figure}[h]
\begin{center}
\includegraphics[scale=0.45]{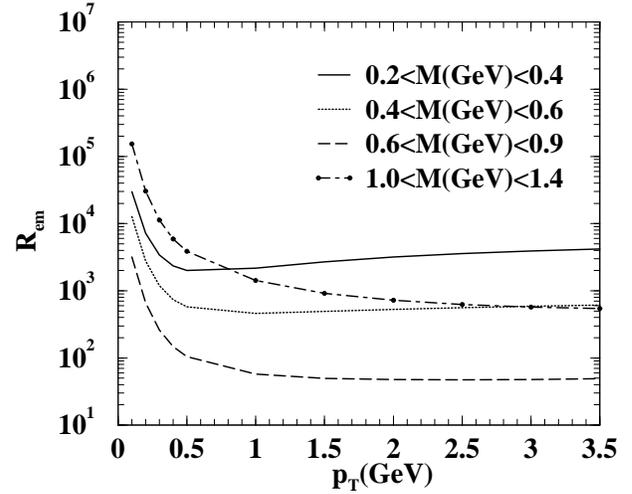}
\caption{Variation thermal photon to dilepton ratio, $R_{em}$ with $p_T$ 
for different invariant mass windows at SPS energy (see text).}
\label{fig5}
\end{center}
\end{figure} 
\begin{figure}[h]
\begin{center}
\includegraphics[scale=0.45]{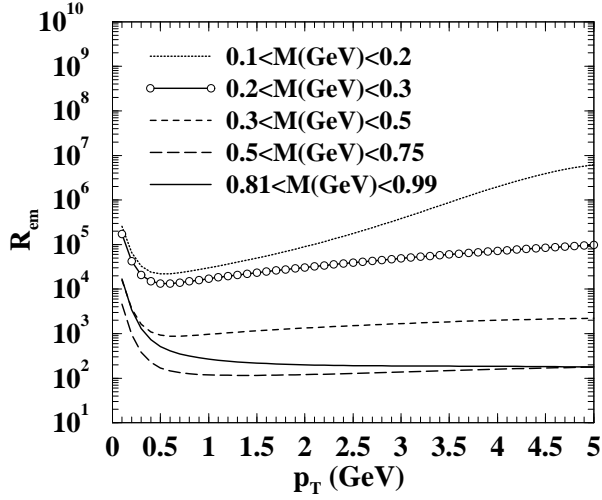}
\caption{Variation thermal photon to dilepton ratio, $R_{em}$ with $p_T$ 
for different invariant mass windows at RHIC energy (see text).}
\label{fig6}
\end{center}
\end{figure} 
\subsection{The ratio, $R_{\mathrm em}$}
As mentioned before some of the uncertainties prevailing in the
individual spectra may be removed by taking the ratio, $R_{\mathrm em}$
of the $p_T$ distribution of thermal photon to dileptons. 
In the absence of experimental data for both photon 
and dilepton from the same colliding system for SPS
energies, we have calculated the ratio $R_{\mathrm em}$ 
for Pb+Pb system, where the initial condition, the freeze-out condition
and the EoS are constrained by the measured WA98 photon spectra. The 
results are displayed in Fig.~\ref{fig5}. 

Next we evaluate the ratio of the 
thermal photon to dilepton spectra constrained
by the experimental data from
Au+Au collisions measured by PHENIX collaboration.
The results for the thermal ratio, $R_{\mathrm em}$  displayed in 
Fig.~\ref{fig6} is constrained by the experimental data.
The behaviour of $R_{\mathrm em}$ with $p_T$ for different invariant 
mass windows which is extracted from the available 
data is similar to the theoretical 
results obtained in Ref.~\cite{JKN1}.
\begin{figure}[h]
\begin{center}
\includegraphics[scale=0.45]{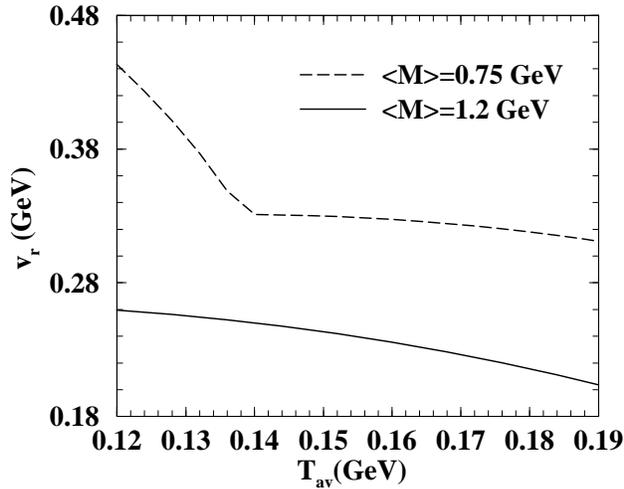}
\caption{The variation of radial flow velocity with 
average temperature of the system for$<M>=0.75$ GeV 
and 1.2 GeV at SPS energy.}
\label{fig7}
\end{center}
\end{figure} 
\begin{figure}[h]
\begin{center}
\includegraphics[scale=0.45]{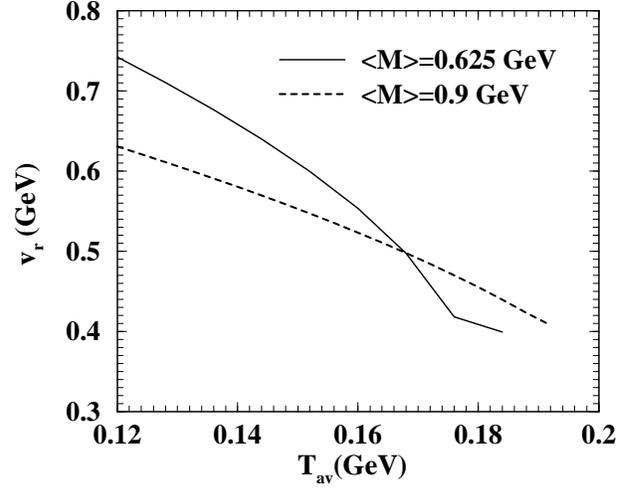}
\caption{The variation of radial flow velocity with average
temperature of the system for$ <M>$ =0.625 GeV and 
0.9 GeV at RHIC energy.}
\label{fig8}
\end{center}
\end{figure} 
The ratio, $R_{em}$ for different $M$ windows (Figs.~\ref{fig5}
and ~\ref{fig6}) can be parametrized as follows:
\begin{equation}
R_{em}=A\left(\frac{M_T}{p_T}\right)^B\exp\left(\frac{M_T-p_T}{T_{eff}}\right)
\label{ratpat}
\end{equation}
with different values of  $T_{eff}$ for 
different invariant mass windows.  The argument of the 
exponential in Eq.~\ref{ratpat} can be written as~\cite{JKN2};  
\begin{eqnarray}
\frac{M_T-p_T}{T_{eff}}&&=\frac{M_T}{T_{eff2}}-\frac{p_T}{T_{eff1}}\nonumber\\
&&=\frac{M_T}{T_{av}+M v_r^2}-\frac{p_T}{T_{av}\sqrt{\frac{1+v_r}{1-v_r}}}
\label{radvel}
\end{eqnarray}
where, $T_{eff1}=T_{av}\sqrt{\frac{1+v_r}{1-v_r}}$ is the blue shifted 
effective temperature for massless photons and $T_{eff2}=T_{av}+M v_r^2$, 
is the effective temperature for massive dileptons. $T_{av}$ is the 
average temperature and $v_r$ is the average radial flow of the system. 
For a given $p_T$ and  $M$ Eq.~\ref{radvel} can be written
as  $v_r=f(T_{av})$. The $T_{eff}$ 
obtained from the parametrization of ratio at SPS energy are 263 MeV 
and 243 MeV for M=0.75 and 1.2 GeV respectively. The average flow 
velocity $v_r$  versus $T_{av}$ have been displayed for M=0.75 GeV and 
1.2 GeV in Fig.~\ref{fig7}. 
The hadronic matter (QGP) dominates the $M\sim 0.75 (1.2)$ GeV region.
Therefore, these two mass windows are selected to extract the flow
parameters for the respective phases.
The $v_r$ increases with decreasing $T_{\mathrm av}$ (increase in time)
and reaches its maximum when the temperature of the system 
is minimum, 
i.e., when the system attains $T_f$, the freeze-out temperature. 
Therefore,  the variation of $v_r$ with $T_{\mathrm av}$ may be treated 
as to show how the flow develops in the system.
The $v_r$ is larger in the hadronic phase because the velocity
of sound in this phase is smaller, which makes 
the expansion slower as a consequence  system lives longer - allowing
the flow to fully develop. On the other hand, $v_r$ is smaller in the
QGP phase because it has smaller life time where the flow is only partially
developed. In Fig.~\ref{fig8} the variation of average transverse 
velocity with average temperature for RHIC initial conditions is depicted. 
The magnitude of the  flow is larger in case of RHIC than SPS 
because of the higher initial pressure. Because of the larger initial
pressure and QGP life time the radial velocity for QGP at RHIC is 
larger compared to SPS.

\par
\begin{figure}[h]
\begin{center}
\includegraphics[scale=0.42]{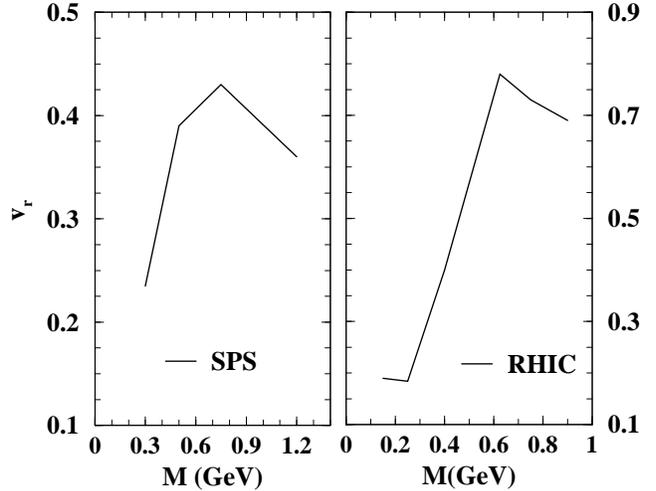}
\caption{The variation of radial flow with invariant mass pairs for 
SPS (left) and RHIC (right) energies.}
\label{fig9}
\end{center}
\end{figure}

 
Obtaining $T_{\mathrm eff1}$ and $T_{\mathrm eff2}$ from the
individual spectra and eliminating $T_{\mathrm av}$ one 
gets the variation of $v_r$ with $M$.
Fig.~\ref{fig9} (left panel) shows the variation of $v_r$ 
with $M$ for SPS conditions. The radial flow velocity 
increases with invariant mass M up to $M=M_{\rho}$ then drops. 
How can we understand this behaviour?
From the invariant mass spectra it is well known that the low 
$M$ (below $\rho$ mass)  and  
high M (above $\phi$ peak) pairs originate from a partonic source~\cite{JKN1}.  
The collectivity (or flow) does not develop fully in the 
QGP because of the small life time of this phase.
Which means that the radial velocity in QGP will be smaller
for both low and high $M$. Whereas the lepton pairs 
with mass around $\rho$-peak mainly originate from a hadronic source 
(at a late stage of the evolution of system) are largely affected 
by the flow resulting in higher values of flow velocity.  
In summary, the value of $v_r$ for $M$ below and above the
$\rho$-peak is small but around the $\rho$ peak is large
- with the resulting behaviour displayed in Fig.~\ref{fig9}.
Similar non-monotonic behaviour is observed in case of
elliptic flow of photon as a function of $p_T$~\cite{v2photon}.
The variation of $v_r$ with $M$ in RHIC (Fig.~\ref{fig9} right panel) 
is similar to SPS,
though the values of $v_r$ at RHIC is larger than that of SPS as expected
due to higher initial pressure.
\section{Summary and Conclusions}
The photon and dilepton spectra measured at SPS and RHIC energies 
by different experimental collaborations have been studied. 
The initial conditions 
have been constrained to reproduce the measured multiplicity in these
collisions. The EoS, the other crucial input to the calculations has 
been taken from lattice QCD calculations.  The deviation of the hadronic
phase from chemical equilibrium is taken in to account by introducing
non-zero chemical potential for each hadronic species.
It is shown that simultaneous
measurements of photon and dilepton spectra in heavy ion collisions
will enable us to quantify the evolution of the average radial flow  
velocity for the system and the nature of the variation of radial flow with 
invariant mass indicate the formation of partonic phase at SPS and RHIC energy.  

{\bf Acknowledgment:} We are grateful to Tetsufumi Hirano for providing
us the hadronic chemical potentials. This work is supported by DAE-BRNS 
project Sanction No.  2005/21/5-BRNS/2455.

\end{document}